\documentclass[12pt]{article}
\usepackage{graphicx}
\usepackage[english]{babel}
\usepackage{cite}
\usepackage{amssymb,amsbsy,amsmath}
\usepackage{calrsfs}

\newcommand{\ha}{\mbox{\small$\frac{1}{2}$}}

\newcommand{\im}{\,{\rm i}\,}
\newcommand{\lab}[1]{\label{#1}}
\newcommand{\re}[1]{(\ref{#1})}

\newcommand{\B}[1]{\mathbf{#1}}

\newcommand{\tG}{{\tilde{G}}}
\newcommand{\D}[2]{{\rm d}^{#1}{#2}\,}
\newcommand{\inta}{\hspace*{-.1cm}\int \hspace*{-.15cm}}

\newcommand{\mc}{\mathrsfs{M}}

\begin{document}

\title{Bound states in the tachyon exchange potential}
\author{Iryna Zahladko$^*$, Askold Duviryak$^\dag$\\
Institute for Condensed Matter Physics\\
of the National Academy of Sciences of Ukraine, Lviv, UA-79011
Ukraine \\$^*$zagladko@icmp.lviv.ua, $^\dag$duviryak@icmp.lviv.ua}

\maketitle

\begin{abstract}
The Klein-Gordon field of imaginary mass is considered as a mediator
of particle interaction. The static tachyon exchange potential is
derived and its applied meaning is discussed. The Schr\"odinger
equation with this potential is studied by means of variational and
numerical methods. Conditions
for existence of bound states are analyzed.
\end{abstract}


\section{Introduction}

The concept of faster-than-light particles -- {\em tachyons} -- is
known more then a half century \cite{BDS62,B-S69,R-M74,Rec86}.
Difficulties in the quantum field theory of tachyons
\cite{Fei67,K-K71} give rise to the idea that free tachyon quanta
cannot exist in nature what is in agreement with experiments
\cite{Rec09}.

Nowadays, the tachyon field is rarely considered as a new kind of
fundamental matter. Rather, it can serve for effective description
of more or less conventional matter in unstable or metastable
states. Effective tachyon fields arise within the quantum gravity
\cite{G-P82,Reb91}, the supersymmetric field theory \cite{BBR95},
the string theory \cite{Mar02,Sen05} etc.

There is a complementary view of tachyons as hidden or virtual
objects \cite{Sud70,Jue73}. Virtual tachyon states (whatever they
are) could participate in an interaction between {\em bradyons}
(i.e., slower-than-light particles) \cite{Spu73,M-R80}. In
particular, peaks in the meson-nucleon differential cross-sections
are effectively treated in \cite{GGSP72} as tachyon resonances.

In the present paper we focus on bound states of particles
interacting via tachyon field which is, in the simplest version, the
Klein-Gordon field with imaginary mass. Once free tachyons are
absent, the symmetric Green function of this field \cite{K-K71} is
an appropriate tool for a description of the tachyon exchange
interaction.

In Section 2 we derive the non-relativistic potential of the tachyon
exchange interaction. An applied treatment of this potential is
discussed. The Schr\"odinger equation with tachyon exchange
potential is solved by means of both the variational approximation
and the numeral integration in Section 3. Conditions for the
existence of bound states are analyzed.


\section{A potential of the tachyon exchange interaction}

One can introduce the tachyon exchange potential in different ways.
First of all, there are known in the literature several
quantum-field descriptions of tachyons \cite{Fei67,K-K71}. Each of
them has some drawbacks or inconsistencies, e.g.
Poincar\'e-non-invariance or/and non-unitarity of S-matrix etc.
These problems are not essential when considering the tachyon
exchange interaction of bradyons in the ladder approximation while
free tachyons are absent. On the second way one can proceed from the
classical action-at-a-distance theory of Wheeler-Feynman type
\cite{W-F45,W-F49,H-N74,H-N95,G-T80,Tre11} in which the
electromagnetic interaction is replaced by the tachyon one. The
third possibility is the partially reduced quantum field theory
\cite{B-D98,D-D00,Dar00,D-D04,ERD05,Z-D12} which takes advantages of
both above approaches. Within this framework quantized matter fields
interact via mediating field (the tachyon one in present case),
variables of which are eliminated from the description at the
classical level.

In all these cases the tachyon exchange interaction of bradyons can
be formulated in terms of the symmetric Green function of tachyon
field, i.e., of the Klein-Gordon field of imaginary mass
$\mc=\im\mu$ \cite{K-K71}, which is real and Poincare-invariant:
%
\begin{eqnarray}\lab{1}
G(x)=\int \frac{\D4k}{(2\pi)^4}\, e^{ik\cdot x}\widetilde{G}(k)
\qquad\mbox{with}\quad \widetilde{G}(k)=-\frac{\cal P}{\mu^2+k^2},
\end{eqnarray}
$x=\{x^{0}, \B x\}$, $k=\{k_{0}, \B k\}$ and ${\cal P}$ standing for
the Cauchy principal value. We use the system of units in which
$c=\hbar=1$ and refer to $\mu$ as \emph{metamass} (following
\cite{BDS62,B-S69}).

Using the Green function of a mediating field one can obtain the
static potential of interaction \cite{G-T80,Tre11,D-D04}. For the
Green function \re{1} we have :
%
\begin{equation}\lab{2}
U(r)=-4\pi\alpha\inta\D{}{x^0}\,G(x-x')=-4\pi\alpha\inta\frac{\D3k}{(2\pi)^3}\mathrm{e}^{-\im{\bf
k\cdot r}}\tG(k_0\!=\!0,\,\B k)=\alpha\frac{\cos{\mu r}}r,
\end{equation}
where $r=|\B r|\equiv|\B x-\B x'|$ is an inter-particle distance and
$\alpha$ is a coupling constant. The static potential is sufficient
for the non-relativistic description of the two-body problem.

Obviously, the tachyon exchange potential \re{2} is equal to the
(real part of) Yukawa potential with the imaginary mass $\mc=\im\mu$
of the mediating field. Alternatively, one can proceed from {\em a
priori} nonrelativistic shielding potential in plasma, substituting
formally the imaginary Debye radius $r_{\rm D}=\im/\mu$. Physically,
this {\em anti-shielding} potential may occur in some metastable
media such as a dielectric at negative temperature \cite{Shen03}.
Another, astrophysical application is the gravity ``dressed'' by a
dark matter \cite{C-T00}.

Here we are interested in the following question: could the
potential \re{2} yield bound states of particles, and, if yes, with
which parameters?


\section{The Schr\"odinger equation}

We suppose that the coupling constant $\alpha$ in \re{2} can be
positive or negative. In both cases the potential \re{2} is the
succession of potential wells and barriers which replace each others
under the change of sign of $\alpha$. Therefore within the classical
(non-quantum) consideration bound states are possible with an
arbitrary $\alpha$ (either negative or positive); they correspond to
a motion of particles within one or another potential well.

The quantum two-body problem is based on the Schr\"odinger equation:
%
\begin{eqnarray}\lab{3}
\frac1{2m}\Delta\Psi(\B r)+ U(r)\Psi(\B r)= {\cal E}\Psi(\B r),
\end{eqnarray}
where $m=m_1m_2/(m_1+m_2)$ is the reduced mass for particle masses
$m_1$, $m_2$, and ${\cal E}$ is the eigen-energy of the system.

By the quantum consideration the depthes of wells or the heights of
barriers of the potential \re{2} become important in view of the
tunneling probability between them. Consequently, properties of the
wave function $\Psi(\B r)$ are not obvious. Nevertheless one can
expect the existence of bound states for $\alpha<0$ and $\mu$ small
when the potential \re{2} is close to the Coulomb one.

In order to solve the Schr\"odinger equation \re{3} we split
variables into the radial and angular ones: $\Psi(\B
r)=\frac1r\psi(r)Y_\ell^\mu(\hat r)$, where $Y_\ell^\mu(\hat r)$ is
the spherical harmonics, and $\hat r=\B r/r$. It is convenient to
introduce the dimensionless variables:
%
\begin{eqnarray}
&&\rho=r/a_{\rm B}, \qquad \epsilon=\mathcal{E}/\mathcal{E}_{\rm
Ry}, \qquad \eta=a_{\rm B}\mu,
\label{4}\\
&&\mbox{where}\qquad a_{\rm B}=1/(m|\alpha|), \qquad
\mathcal{E}_{\rm Ry}=m\alpha^2 \qquad \label{5}
\end{eqnarray}
are analogs of the Bohr radius and the Rydberg constant. Then the
equation for the radial wave function $\psi(\rho)$ takes the form:
%
\begin{eqnarray}
&&\{H^{(\mp)}-\epsilon\}\psi(\rho)=0, \label{6}\\
\mbox{where}&&
H^{(\mp)}\equiv\frac12\left\{-\frac{\mathrm{d}^2}{\mathrm{d}\rho^2} +
\frac{\ell(\ell+1)}{\rho^2}\right\}+u^{(\mp)}(\rho),
\label{7}\\
&&u^{(\mp)}(\rho)=\mp\frac{\cos\eta\rho}{\rho}
\equiv\mp\frac1{2\rho}\left\{\mathrm{e}^{\im\eta\rho}+\mathrm{e}^{-\im\eta\rho}\right\}.
\label{8}
\end{eqnarray}
The operators $H^{(\mp)}$ and $u^{(\mp)}$ correspond to the cases
$\alpha=\mp|\alpha|\lessgtr0$ which we will conventionally refer to
as the "attraction" and "repulsion" ones. We note that these terms
are rather provisional and they do not reflect actual properties of
the interaction.

\subsection*{The variational solution}

There is unknown the exact solution of the  Schr\"odinger equation
with the potential \re{8} which is the superposition of Yukawa
potentials with imaginary exponents. We expect that the calculation
of the ground state energy by means of the variational Ritz method
may appear efficient, by analogy to the case of ordinary Yukawa
potential \cite{Flu74}.

Since the potential \re{8} is close to the Coulomb potential at
$\rho\ll1/\eta$, we choose as a trial wave function the following
simple one:
%
\begin{equation}\label{9}
\tilde\psi_{\ell0}\equiv\sqrt{k}\psi_{\ell0}(k\rho);
\end{equation}
here $k$ is the variational (scaling) parameter, and the function
$\psi_{\ell0}(\rho)$ describes the ground state ($n_r=0$) of the
radial Hamiltonian $H^{(-)}_{\eta=0}$ \re{7}-\re{8} of the Coulomb
problem \cite{L-L63}. The function \re{9} is normalized and properly
behaved at $\rho\to0$.

Using the integration techniques with hypergeometric functions
\cite{L-L63} we obtain the following expression for the average
energy on the variational state \re{9}:
%
\begin{eqnarray}
\langle\epsilon\rangle_\varkappa\equiv\left\langle-\frac12\frac{\D{2}{}}{\D{}{\rho^2}}+
\frac{\ell(\ell+1)}{2\rho^2}+u(\rho)\right\rangle_\varkappa
&=&\frac{\eta^2}{8\varkappa^2}
- \frac\eta{4n\varkappa}f(\varkappa), \label{10}\\
\mbox{where}\quad
f(\varkappa)&=&\frac1{(1+\im\varkappa)^{2n}}+\frac1{(1-\im\varkappa)^{2n}},
\label{11}
\end{eqnarray}
$n=\ell+1$, and the new variational parameter $\varkappa=n\eta/(2k)$
(instead of $k$) is used.

The minimum condition
$\frac{\partial}{\partial\varkappa}\langle\epsilon\rangle_\varkappa=0$
for the function \re{10} yields the equality:
%
\begin{equation}\label{12}
\eta=\frac{\varkappa}{n}\{f(\varkappa)-\varkappa f'(\varkappa)\}
\end{equation}
which, together with \re{10}, determines the ground state energy as
a parametric function (in terms of the parameter $\varkappa$) of the
dimensionless metamass $\eta$.

Let us consider the ground s-state by putting $n=1$. It follows from
the equations \re{10}-\re{12} that $\langle\epsilon\rangle\le0$
while $\varkappa\in[0,\sqrt2{-}1]$. Hereby the energy
$\langle\epsilon\rangle\in[-\ha,0]$ grows up monotonously together
with the dimensionless metamass $\eta\in[0,1]$. Since
$\langle\epsilon\rangle$ exceeds the true values of energy, the
bound state certainly exists for $\eta\in[0,1]$. Thus we can
estimate the critical value $\eta_{\rm c}$ of metamass (at which
bound states disappear) from below as $\eta_{\rm c}\ge1$.

An analytical method was developed by Sedov \cite{Sed70} for
determining the critical screening parameter for the Yukawa
potential $v(\rho)=-\mathrm{exp}(-\eta\rho)/\rho$. It was then
applied to the related exponential cosine screened Coulomb (ECSC)
potential \cite{Dut80}. The method leads to the equation:
%
\begin{equation}\label{13}
1-I_1/\eta_{\rm c}+I_2/\eta_{\rm c}^2 -I_3/\eta_{\rm c}^3
+I_4/\eta_{\rm c}^4-\dots=0,
\end{equation}
where the coefficients $I_n$ ($n=1,2,\dots$) can be expressed via
multiple quadratures. In the present case \re{8} a direct
application of the method fails since a part of quadratures
diverges. We consider, instead of \re{8}, the extended ECSC
potential $w(\rho)=\mathrm{e}^{-\xi\rho}u^{(\mp)}(\rho)$, $\xi>0$,
and calculate asymptotic values of quadratures $I_n$ at $\xi\to0$.
We obtain $I_1\sim\pm2\xi$, $I_2\sim2\ln\xi$, $I_3\to\mp\pi$ and
$I_4\sim2\ln^2\xi$ but an evaluation of higher-order coefficients is
an overpowering task. In 2nd- and 3rd-order approximations the
equation \re{13} possesses the real solution $\eta_{\rm c}
\sim\sqrt{-2\ln\xi}\to\infty$ but in 4th order such the infinite
solution turns into a complex number and thus it cannot be a
critical value. Nothing is known about an exact solution or
higher-order solutions.

\subsection*{The numeral solution}

In principle, the true values of energy can be determined with an
arbitrary precision by numeral integrating the Schr\"odinger
equation \re{6}. Figure 1a shows the dependency $\epsilon(\eta)$
which is obtained in both the variational and numeral ways, for the
Hamiltonian $H^{(-)}$, i. e. for the "attraction" case $\alpha<0$.
It is obvious the variational approximation is satisfactory all over
the segment $\eta\in[0,1[$, except for $\eta\gtrsim1$ where it is
not correct. On the other hand, numeral integration yields bound
states up to $\eta\approx2.5$. Around this value the binding energy
becomes negligibly small: $|\epsilon|_{\eta>2.5}<10^{-9}$, and the
technical difficulties of the numeral integration grow up quickly.
Thus, in such a way we cannot determine the critical value of
metamass $\eta_{\rm c}$ (if finite) at which bound states cease.

It is obvious from Figure 1b that the dependency of
$1/\ln|\epsilon|$ on $\eta$ tends to the asymptote with the abscissa
intercept about $3$. Thus bound states credibly are absent for
$\eta>\eta_{\rm c}\approx3$. Of course, this assumption requires a
rigorous proof  which we failed to provide. In practice, an exact
value of $\eta_{\rm c}$ is not crucially important since states with
negligibly small binding energy may be considered as unbound.

\subsection*{The "repulsion" case}

The variational approximation of bound states for the "repulsion"
case $\alpha>0$ can be obtained from the equations \re{10}-\re{12}
as well if one puts formally $\eta<0$ and considers that $|\eta|$ is
the dimensionless metamass (instead of $\eta$). In this case
$\langle\epsilon\rangle\le0$ for $\eta\in[0,-1]$. Thus the existence
of bound states in the "repulsion" case has been proven too in
despite that the used variational ansatz \re{9} appears extremely
crude. Numerical results in figure 2a show that bound states exist
for $0<|\eta|\le2.7$ and probably do not exist for $|\eta|\ge3$
(similarly to the case $\alpha<0$); see figure 2b. The maximum of
the binding energy $|\epsilon|_{\rm max}\approx0.0764$ (reached at
$|\eta|\approx3/4$) is almost to the one degree smaller than one in
the case $\alpha<0$. This ability of the tachyon interaction to bind
particles in both cases of negative and positive coupling constant
(but with different intensions) is distinctive.
%
\begin{figure}[hb]
\includegraphics[scale=0.9]{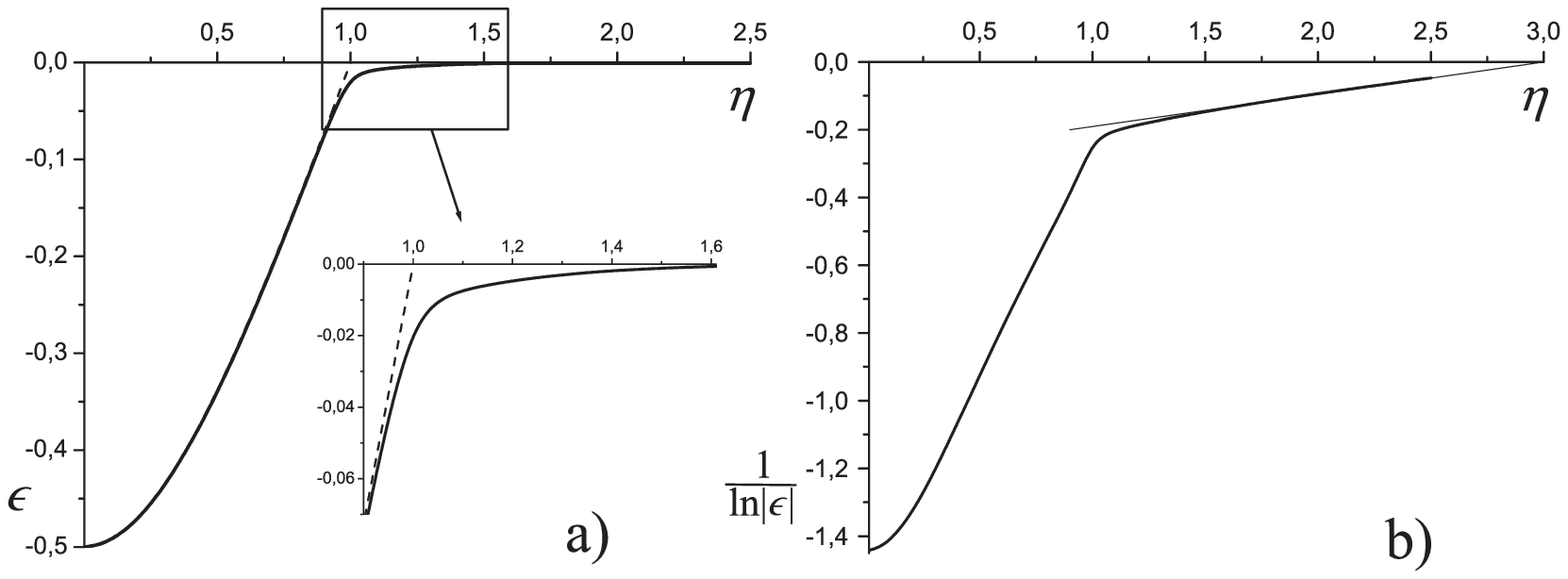}
\caption{\small{\large a)} the dependency of the ground state energy
on the metamass of the tachyon mediator of interaction (in the
dimensionless variables $\epsilon$, $\eta$) for the "attraction"
case $\alpha<0$: solid line -- numerical results, dashed line --
variational approximation; {\large b)} asymptotical behavior of the
energy for $1<\eta<3$.} \bigskip
\includegraphics[scale=0.9]{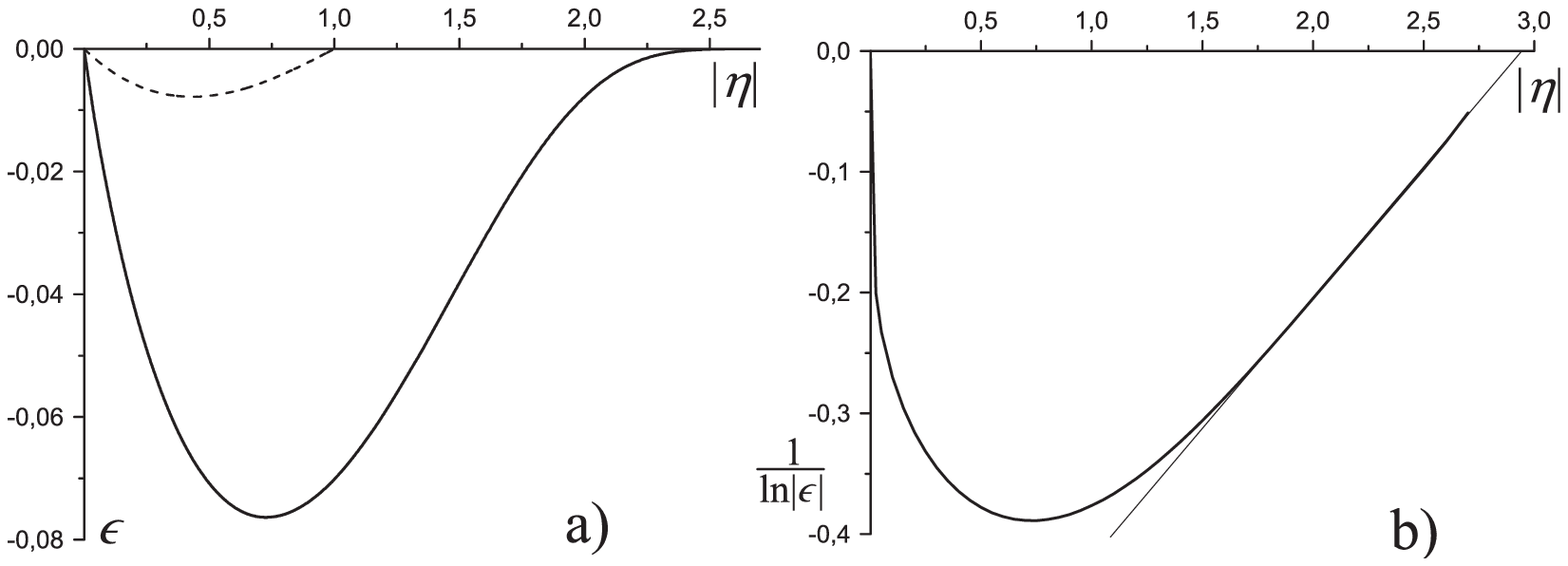}
\caption{\small The same as in Figure 1, but for the "repulsion"
case $\alpha>0$.}
\end{figure}


\section{Conclusions}

In the two-particle problem with the tachyon exchange interaction we
limit ourselves by the non-relativistic consideration and analyze
the Schr\"odinger equation with the anti-shielding potential, i.e.,
the Coulomb potential modulated by $2\pi/\mu$ periodical cosine
\re{2}.

Since the exact solution of this equation is unknown, the
variational and numeral methods were applied. With the former it is
proved that bound states exist at both the negative and positive
coupling constant $\alpha$ for the metamass $\mu=\eta\,m|\alpha|$ in
the range $0<\eta<1$. The numeral calculations indicate the
existence of bound states for $1\le\eta\le2.5$ too. Moreover, the
ground state binding energy goes down quickly by increase of the
metamass such that $|{\cal E}/{\cal E}_{\rm Ry}|<10^{-9}$ at
$\eta=2.5$. Numeral integration for $\eta>2.5$ becomes overwhelming
and do not provide reliable results. The extrapolation of the
dependency of ${\cal E}$ on the metamass in the area $\eta>2.5$
suggests further decrease of the binding energy up to zero at
$\eta_{\rm c}\simeq3$, or $\mu_{\rm c}\simeq3m|\alpha|$. Thus bound
states exist for $\mu<\mu_{\rm c}$ or, in terms of imaginary Debye
radius, for $|r_{\rm D}|>a_{\rm B}/3$. The Sedov method
\cite{Sed70,Dut80} does not work in the present case, thus an exact
or, at least, a precise (if finite) value of the critical parameter
$\eta_{\rm c}$ is an open question.



\begin{thebibliography}{99}
\bibitem{BDS62}
O.-M. Bilaniuk, V. K. Deshpande and E. C. G. Subarshan,  Amer. J.
Phys. 30, 718 (1962).
\bibitem{B-S69}
O.-M. Bilaniuk, V. K. Deshpande and E. C. G. Subarshan, Physics
Today 22, 43 (1969).
\bibitem{R-M74}
E. Recami and R. Mignani, Riv. Nuovo Cim. 4, 209 (1974).
\bibitem{Rec86}
E. Recami,  Riv. Nuovo Cim. 9, 1 (1986).
\bibitem{Fei67}
G. Feinberg, Phys. Rev. 159, 1089 (1967).
\bibitem{K-K71}
K. Kamoi and S. Kamefuchi, Prog. Theor. Phys. 45, 1646 (1971).
\bibitem{Rec09}
E. Recami,  J. Phys.: Conf. Ser. 196, 012020 (2009).
\bibitem{G-P82}
D. J. Gross and J. Perry, Phys. Rev. D 25, 330 (1982).
\bibitem{Reb91}
A. Rebhan, Nucl. Phys. B 351, 706 (1991).
\bibitem{BBR95}
D. G. Barci, C. G. Bollini and M. C. Rocca, Internat. J. Mod. Phys.
A 10, 1737 (1995).
\bibitem{Mar02}
A. V. Marshakov, Physics--Uspekhi, 45, 977 (2002).
\bibitem{Sen05}
A. Sen, Internat. J. Theor. Phys. 20, 5513 (2005).
\bibitem{Sud70}
E. C. G. Sudarshan, Phys. Rev. D 1, 2428 (1970).
\bibitem{Jue73}
Ch. Jue, Phys. Rev. D 8, 1757 (1973).
\bibitem{Spu73}
E. van der Spuy, Phys. Rev. D 7, 1106 (1973).
\bibitem{M-R80}
G. D. Maccarrone and E. Recami, Nuovo Cim. A 57, 85 (1980).
\bibitem{GGSP72}
A. M. Gleeson, M. G. Gundzik, E. C. G. Sudarshan and A. Pagnamenta,
Phys. Rev. D 6, 807 ( 1972).
\bibitem{W-F45}
J. A. Wheeler and R. P. Feynman, Rev. Mod. Phys. 17, 157 (1945).
\bibitem{W-F49}
J. A. Wheeler and R. P. Feynman, Rev. Mod. Phys. 21, 425 (1949).
\bibitem{H-N74}
F. Hoyle and J. V. Narilikar, Action at a distance in physics and
cosmology (New York, Freemen, 1974).
\bibitem{H-N95}
F. Hoyle and J. V. Narilikar, Rev. Mod. Phys. 67, 113 (1995).
\bibitem{G-T80}
R. P. Gaida and V. I. Tretyak, Acta Phys. Pol. B 11, 502 (1980).
\bibitem{Tre11}
V. I. Tretyak, Forms of relativistic Lagrangian dynamics
(Ky{\"{\i}}v, Naukova Dumka, 2011) [in Ukrainian].
\bibitem{B-D98}
M. Barham and J. W. Darewych, J. Phys. A 31, 3481 (1998).
\bibitem{D-D00}
B. Ding and J. Darewych, J. Phys. G  26, 907 (2000).
\bibitem{Dar00}
J. Darewych,  Condens. Matter Phys. 3, 633 (2000).
\bibitem{D-D04}
A. Duviryak and J. W. Darewych,  J. Phys. A 37, 8365 (2004).
\bibitem{ERD05}
M. Emami-Razavi and J. W. Darewych, J. Phys. G  31, 1095 (2005).
\bibitem{Z-D12}
I. Zagladko and A. Duviryak, J. Phys. Studies 16, 3101 (2012).
\bibitem{Shen03}
J.-Q. Shen, Phys. Scripta 68, 87 (2003).
\bibitem{C-T00}
T. Chiueh and Y.-H. Tseng, Astrophys. J. 544, 204 (2000).
\bibitem{Flu74}
S. Fl\"ugge, Practical quantum mechanics I (Berlin-Heidelberg-NY,
Springer-Verlag, 1971).
\bibitem{L-L63}
L. Landau and E. Lifshitz, Quantum Mechanics: Non-Relativistic
Theory, Vol. 3 (Oxford, Pergamon Press, 1981).
\bibitem{Sed70}
V. L. Sedov. Sov. Phys. Dokl. 14, 871 (1970).
\bibitem{Dut80}
R. Dutt, Phys. Lett. A 77, 229 (1980).
\end{thebibliography}
\end{document}